\documentclass[conference]{IEEEtran}
\IEEEoverridecommandlockouts
\usepackage{amsmath,graphicx}
\usepackage{url}
\usepackage{hyperref}
\usepackage{subcaption}
\usepackage{url}
\usepackage{multicol}
\usepackage{multirow}
\usepackage{mdwlist}
\usepackage{enumitem}
\usepackage{amsthm}
\usepackage{booktabs}
\usepackage{algorithm}
\usepackage{algorithmic}
\usepackage[switch]{lineno}
\usepackage{tabularx}
\usepackage{amssymb}
\usepackage{xcolor}
\usepackage{cite}

\def\BibTeX{{\rm B\kern-.05em{\sc i\kern-.025em b}\kern-.08em
    T\kern-.1667em\lower.7ex\hbox{E}\kern-.125emX}}

\makeatletter
\renewcommand{\maketag@@@}[1]{\hbox{\m@th\normalsize\normalfont#1}}%
\makeatother
\newcommand{\para}[1]{{\vspace{1.5pt} \bf \noindent #1 \hspace{4pt}}}

\begin{document}
%\ninept
%
% Title.
% ------
\title{CycleFlow: Leveraging Cycle Consistency in Flow Matching for Speaker Style Adaptation}

\author{
    \IEEEauthorblockN{Ziqi Liang$^{2,3\dagger}$, Xulong Zhang$^{1\dagger}$\thanks{$\dagger$ Equal Contributions}, Chang Liu$^{2}$, Xiaoyang Qu$^{1}$, Weifeng Zhao$^{3}$, Jianzong Wang$^{1\ast}$\thanks{$^\ast$Corresponding author: Jianzong Wang (jzwang@188.com)}}
    \IEEEauthorblockA{
        \textit{$^{1}$ Ping An Technology (Shenzhen) Co., Ltd., Shenzhen, China}\\
        \textit{$^{2}$ University of Science and Technology of China, Hefei, China}\\
        % \textit{$^{3}$Tencent Music Entertainment Lyra Lab, Shenzhen, China}\\
        \textit{$^{3}$ Tencent Lyra Lab, Shenzhen, China}\\
    }
}

% \author{
%     \IEEEauthorblockN{Ziqi Liang$^{2,3\dagger}$, Xulong Zhang$^{1\dagger}$\thanks{$\dagger$ Equal Contributions}, Chang Liu$^{2}$, Xiaoyang Qu$^{1}$, Weifeng Zhao$^{3}$, Jianzong Wang$^{1\ast}$\thanks{$^\ast$Corresponding author: Jianzong Wang (jzwang@188.com)}}
%     \IEEEauthorblockA{
%         \textit{$^{1}$Ping An Technology (Shenzhen) Co., Ltd., Shenzhen, China}\\
%         \textit{$^{2}$University of Science and Technology of China, Hefei, China}\\
%         \textit{$^{3}$Tencent Music Entertainment Lyra Lab, Shenzhen, China}\\
%     }
% }

% \author{
%     \IEEEauthorblockN{Ziqi Liang$^{1,2\dagger}$, Xulong Zhang$^{1\dagger}$\thanks{$\dagger$ Equal Contributions}, Chang Liu$^{2}$, Xiaoyang Qu$^{1}$, Jianzong Wang$^{1\ast}$\thanks{$^\ast$Corresponding author: Jianzong Wang (jzwang@188.com)}}
%     \IEEEauthorblockA{
%         \textit{$^{1}$Ping An Technology (Shenzhen) Co., Ltd., Shenzhen, China}\\
%         \textit{$^{2}$University of Science and Technology of China, Hefei, China}
%     }
% }

\maketitle
\begin{abstract}
Voice Conversion (VC) aims to convert the style of a source speaker, such as timbre and pitch, to the style of any target speaker while preserving the linguistic content. 
However, the ground truth of the converted speech does not exist in a non-parallel VC scenario, which induces the train-inference mismatch problem.
Moreover, existing methods still have an inaccurate pitch and low speaker adaptation quality, there is a significant disparity in pitch between the source and target speaker style domains.
% There is a significant disparity in speaker style, such as pitch, between the source and target domains in unseen speaker scenarios. 
As a result, the models tend to generate speech with hoarseness, posing challenges in achieving high-quality voice conversion.
In this study, we propose CycleFlow, a novel VC approach that leverages cycle consistency in conditional flow matching (CFM) for speaker timbre adaptation training on non-parallel data. 
% We can shift our focus from reconstruction to conversion even in the training scenario.
Furthermore, we design a Dual-CFM based on VoiceCFM and PitchCFM to generate speech and improve speaker pitch adaptation quality.
% The former generates F0 with the target speaker style, and the generated F0 is fed to the latter to convert the speech with a target speaker style.
Experiments show that our method can significantly improve speaker similarity, generating natural and higher-quality speech.
% Audio samples can be found at \href{}{}.
\end{abstract}


\section{Introduction}
Voice conversion aims to transfer the speaking style of a source speaker to the speaking style of a target speaker while maintaining the original linguistic content.
It holds significance in various scenarios, such as movie dubbing, live broadcasting, and data augmentation, offering the potential to assist people in expressing themselves in a suitable speaking style. 

In recent years, there has been rapid progress in speech representation distanglement for voice conversion. Many previous approaches~\cite{UUVC,PolyakACKLHMD21,eadvc,NANSY,DengT0W0023} utilize pretrained self-supervised learning (SSL) models~\cite{wav2vec2,hubert,contentvec} or automatic speech recognition (ASR) to extract bottleneck features as linguistic content distangled from the source speech.
Simultaneously, a pre-trained speaker verification model is used to extract speaker representation. Then VC model combines the source linguistic content with target speaker representation to reconstruct the converted speech.
Despite the previous approach \cite{contentvec,CosyVoice,soundstorm,SEF_VC,WangCXTW23,TangZWCX23a} based on discrete tokens achieving some success in content decoupling, there is still a timbre leakage in the extracted content representation, leading to poor speaker simiarity in the converted speech. 

Limited by the lack of a well-decoupled content encoder and non-parallel VC scenario, learning a mapping between two domains with unparallel training samples is a more effective approach.
Some generative adversarial network (GAN)-based methods \cite{cycleganvc,cycleganvc2,maskcycleganvc,ZhangWCXX21} employ cycle consistency to learn this mapping without any parallel data. 
% It requires two mappings $\mathcal{X}$$\to$$\mathcal{Y}$ and $\mathcal{Y}$$\to$$\mathcal{X}$. 
Cycle consistency enforces transitivity between forward and inverse conversion by regularizing pairs of samples, particularly in non-parallel VC where no correspondence between speech in source and target domains is guaranteed.
However, training GAN-based models is typically challenging, as it is unstable and can easily lead to gradient explosion or vanishing.

Moreover, challenges persist in cross-domain VC, where the pitch of the source domain speaker exceeds the vocal range of the target domain speaker. It leads to hoarseness in cases where the pitch span is large, which hinders the applicability of VC in scenarios requiring transformations across extensive vocal ranges. This is common in real-world scenarios, but existing VC models rarely focus on this.

\begin{figure}[t]
    \centering
    \includegraphics[width=8.0cm]{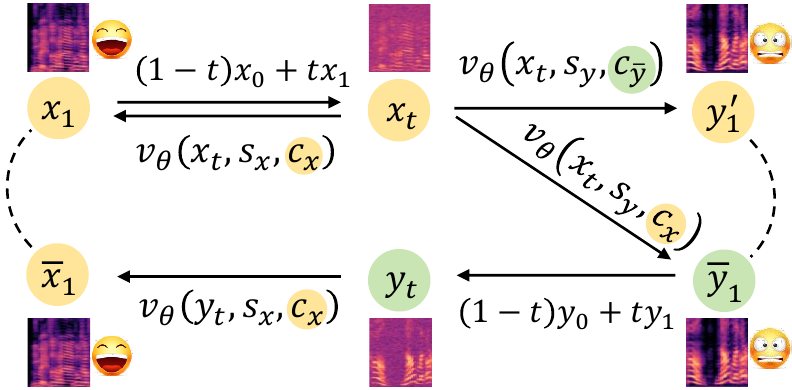}
    % \vspace{-0.2cm}
    \caption{The conversion cycle in CFM-based VC.} 
    \vspace{-0.6cm}
    \label{cycleflow}
\end{figure}

% idea
% 1. cycle regular 实现non-parallel vc for Timbre adaptation
% 2. dual-cfm 实现 pitch 校准Pitch adaptation

In this paper, we propose CycleFlow to improve speaker style adaptation from two aspects: timbre and pitch.
We design cycle consistency regularization to improve speaker timbre adaptation, training on non-parallel data without a well-decoupled content representation.
To improve speaker pitch adaptation, we designed a Dual-CFM for speech generation and pitch correction, which alleviates generated speech's hoarseness in cross-domain VC scenarios caused by large pitch spans.
To summarize, our contributions are as follows:\

\begin{enumerate}
\item We propose cycle consistency regularization to increase speaker timbre similarity training on non-parallel data.
\item We design the Dual-CFM, which contains PitchCFM and VoiceCFM for pitch correction and speech generation. 
\item Experiments demonstrate CycleFlow's superiority in pitch correction and speaker style adaptation.
\end{enumerate}

% 1. 针对cosyvoice的speech token, ASR训的token，合成的音频由于缺乏F0引导信息，存在hoarseness
% 2. 我们使用F0-cycle training, 在不需要额外数据或增加模型参数的前提下提升VC音频的质量
% 3. 

% 它可以在不需要额外数据或增加模型参数的情况下提高SVC任务中的语音质量。我们创新性地在我们的 SVC 模型中引入了一个循环音高移动训练策略和结构相似性指数 (SSIM) 损失，有效地提高了其性能。在公共歌唱数据集 M4Singer 上的实验结果表明，我们提出的方法在一般 SVC 场景中显着提高了模型性能，尤其是在跨域 SVC 场景中。

\begin{figure*}[t]
    \centering
    \includegraphics[width=16.0cm]{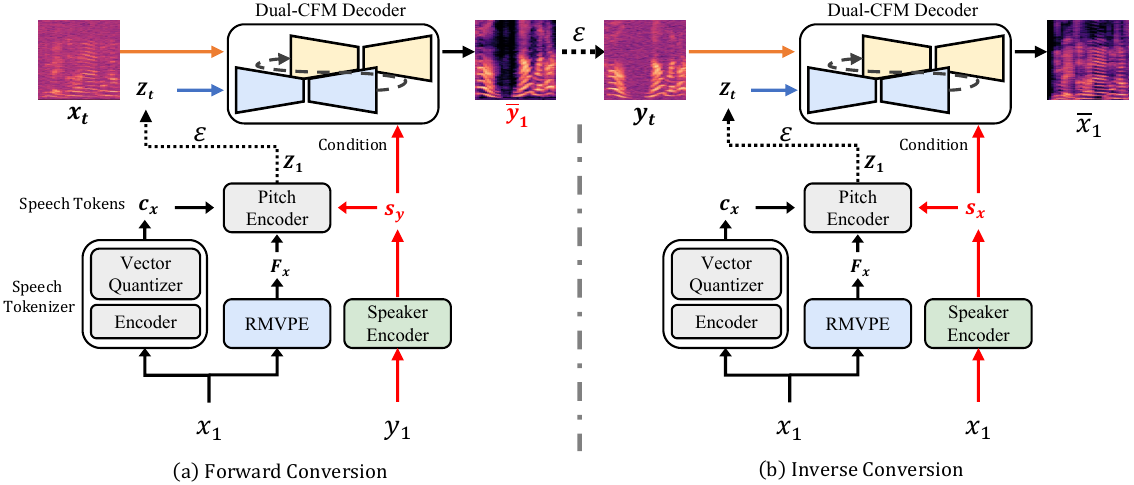}
    \vspace{-0.1cm}
    \caption{ Overall pipeline of proposed CycleFlow. The CycleFlow includes a forward conversion from the source speech $x_1$ to the converted speech $\bar{y}_{1}$, followed by a inverse conversion to the reconstructed source speech $\bar{x}_{1}$.} 
    \vspace{-0.3cm}
    \label{CycleFlow2}
\end{figure*}

\section{Method}

\subsection{Speech disentanglement}
As shown in Fig.~\ref{CycleFlow2}, we first distangle speech into representation of content, pitch, and timbre: 

\para{Content:}Speech tokenizer from \cite{CosyVoice} aims to compress speech into supervised semantic tokens, which are derived from ASR model by inserting vector quantization (VQ) into the encoder's initial six layers. 
% VQ utilizes a solitary codebook with an expansive dictionary containing 4096 entries. The derived token sequence exhibits a frequency of 50Hz.
Since the speech tokenizer is trained to minimize the recognition errors of rich text in an end-to-end manner, the extracted tokens have a strong semantic relationship to linguistic information. 

\para{Pitch:}We utilize the pre-trained RMVPE \cite{WeiCDC23} model to extract the $F_0$, for its effectiveness and robustness.

\para{Timbre:}Speaker embedding vector is extracted from the reference audio with a pre-trained speaker encoder\cite{WangZCC023}.
The speaker embedding serves as a guide for Dual-CFM decoder.

\subsection{Cycle consistency regularization}
As illustrated in Fig.~\ref{cycleflow}, our goal is to enable the CFM model to convert a source speech $x_1$ by using it as the content condition $c_x$, and then denoise the noised latent $y_t$ to $y_1$ with target speaker style $s_y$. To learn such a VC model $v_{\theta}(y_t, s_y, c_x)$, we design the cycle consistency regularization to ensure cycle consistency so that the source semantics and target speaker style are preserved in the converted speech.

% We assume a likelihood function $P(z_0,s_{\mathrm{x}})$ that the image $x_0$ falls into the data distribution specified by the text condition $s_{\mathrm{x}}$.
% We consider a generalized case of cycle consistency given the conditioning mechanism in LDMs.
% If $P(c_{\mathrm{x}},s_{\mathrm{x}})$ is close to 1, i.e., the condition $c_{\mathrm{x}}$ falls exactly into the data distribution described by the speaker style condition $s_{\mathrm{x}}$, we should expect that $G(z_t,c_{\mathrm{text}},c_{\mathrm{img}}) = c_{\mathrm{img}}$ for any noised latent $z_t$.

With the cyclic conversion in Fig.~\ref{cycleflow}, a set of consistency losses is given by dropping time-dependent variances:
% \begin{align}
\begin{equation}
\label{const}
\begin{split}
    \mathcal{L}_{x\rightarrow x} = &\mathbb{E}_{x_0,\varepsilon}\  
    || v_{\theta}(x_t,s_x,c_x) - (x_1-(1-\sigma)x_0) ||_2^2 \\
    \mathcal{L}_{y \rightarrow y} = &\mathbb{E}_{x_0,\varepsilon} \ 
    || v_{\theta}(y_t,s_y,c_{\Bar{y}_0}) - (y_1-(1-\sigma)y_0) ||_2^2 \\
    \mathcal{L}_{x \rightarrow y \rightarrow x} = &\mathbb{E}_{x_0,\varepsilon} \ 
    || v_{\theta}(y_t,s_x,c_x) - (y_1-(1-\sigma)y_0) \\  &\qquad + v_{\theta}(x_t,s_y,c_x) - (x_1-(1-\sigma)x_0)  ||_2^2 \\
    \mathcal{L}_{x \rightarrow y \rightarrow y} = &\mathbb{E}_{x_0,\varepsilon} \ 
    || v_{\theta}(x_t,s_y,c_x) - v_{\theta}(x_t,s_y,c_{\Bar{y}_0}) ||_2^2
\end{split}
\end{equation}
% \end{align}

\para{Reconstruction loss:}$\mathcal{L}_{x\rightarrow x}$ and $\mathcal{L}_{y\rightarrow y}$ ensures that CycleFlow can function as a conditional flow matching to reverse a speech.

\para{Cycle consistency loss:}$\mathcal{L}_{x \rightarrow y \rightarrow x}$ serves as the transitivity regularization, which ensures that the forward and inverse conversion can reconstruct the original speech $x_1$.

\para{Invariance loss:}$\mathcal{L}_{x \rightarrow y \rightarrow y}$ requires that the target speaker style domain stays invariant under forward conversion, i.e., given a forward conversion from $x_t$ to $\Bar{y}_1$ conditioned on $c_x$ and $s_y$, repeating the conversion conditioned on $\Bar{y}_1$'s speaker style $s_y$ and content $c_y$ would reproduce $\Bar{y}_1$. 

The training objectice of conversion is thus given by:
\begin{equation}
\begin{split}   
        \mathcal{L}_x=\lambda_1\mathcal{L}_{x \rightarrow x}+\lambda_2\mathcal{L}_{x \rightarrow y \rightarrow x}+\lambda_3\mathcal{L}_{x \rightarrow y \rightarrow y} 
\end{split}
\end{equation}

Consider both conversion cycle from $x \leftrightarrow y$, the complete training objective of CycleFlow is:
\begin{equation}
\begin{split}   
        \mathcal{L}_{Cycle}=\mathcal{L}_{x}+\mathcal{L}_{y} 
\end{split}
\end{equation}

\subsection{Dual-CFM decoder}

Compared with diffusion-based VC \cite{PopovVGSKW22,Diff-HierVC}, the VC model \cite{soundstorm,ns2} utilizing a speech codec to reconstruct converted speech from speech tokens, resulting in poor speech quality. 
\cite{CosyVoice} employs optimal-transport conditional flow matching (OT-CFM) \cite{LipmanCBNL23,0001FMHZRWB24} to generate speech from speech tokens, which can achieve simpler gradients, easier training and faster generation than diffusion.
Inspired by \cite{CosyVoice,Diff-HierVC}, we design a Dual-CFM decoder which contains PitchCFM and VoiceCFM. PitchCFM initially converts the source $F_0$ with the target speaker style, and the converted $F_0$ is fed to VoiceCFM to convert the speech with target speaker style hierarchically. The details of each CFM models are described as follows. 

\para{PitchCFM}is a pitch generator based on OT-CFM. 
% We employ it to learn the denoising process from noise into the pitch distribution. 
We employ it to learn the transformation $Z_t$ into $F_0$ sample $Z_1$, starting from a standard Gaussian distribution $Z_0$. 
\begin{equation}
\begin{split}   
        % &\frac{d}{dt}Z_{t}=v_{t}(Z_{t},t)\\
        Z_t=(1-(1-\sigma)t)Z_0+tZ_1
\end{split}
\end{equation}
$Z_t$ is governed by an ordinary differential equation (ODE), and it executes denoising to recover the original pitch contour in the reverse process.

We define the optimal-transport (OT) flow and force a neural network $\mathrm{NN}_{pitch}$ to estimate the vector field $v_t$:
\begin{equation}
\begin{split}   
        v_t(Z_{t},t)=\frac{d}{dt}Z_{t}=Z_1-(1-\sigma)Z_0
\end{split}
\end{equation}
Where $\sigma$ is a constant. As illustrated in Fig.~\ref{CycleFlow2}, speech tokens $c$ are first encoded by a Conformer \cite{conformer} and then upsampled to match the leagth of the $F_0$ embedding. The encoded feature $c$, target speaker embedding $s$, and source $F_0$ embedding $f$, are fed into the pitch encoder \cite{Diff-HierVC} to produce the pitch embedding $Z_1$.
The sample $Z_t$ generated from $Z_1$ is fed into UNet-based pitch score estimator for vector field prediction:
\begin{equation}
\begin{split}   
        v_{t}^{\mathrm{pred}}=\mathrm{NN}_{pitch}(Z_t,t;s)
\end{split}
\end{equation}

The pitch encoder \cite{Diff-HierVC} transforms the normalized $F_0$ of the source speaker to the pitch representation $Z_1$. We regularized the pitch representation by pitch reconstruction loss to utilize $Z_1$ as a prior distribution of PitchCFM as follow:
\begin{equation}
\begin{split}   
        \mathcal{L}_{pitch}=\| v_{t}^{pred}-v_t \|^{2}
\end{split}
\end{equation}

Since the speech tokens contains limited timbre information, we use a cross-attention in \cite{conformer} to extract target speaker style from the reference audio.
At inference, the pitch embedding $Z_1$ from the pitch encoder is utilized as a prior of PitchCFM, and it generates the refined $F_0$ with the target speaker style.

\para{VoiceCFM}is a Mel-spectrogram generator based on OT-CFM for high-quality speech synthesis from content, target $F_0$, and target speaker's style. 
It can utilize the PitchCFM's output refined $F_0$ and target speaker style $s$ as condition to maximize speaker adaptation capacity.
We employ it to learn the transformation $x_t$ into the distribution of Mel-spectrogram $x_1$, starting from a standard Gaussian distribution $x_0$. 
\begin{equation}
\begin{split}   
        % &\frac{d}{dt}x_{t}=v_{t}(x_{t},t)\\
        x_t=(&1-(1-\sigma)t)x_0+tx_1
\end{split}
\end{equation}
Where transformation $x_t$ is governed by an ODE, and we also define an OT-flow and force a neural network $\mathrm{NN}_{mel}$ to estimate the vector field $v_t$:
\begin{equation}
\begin{split}   
        v_t(x_{t},t)=\frac{d}{dt}x_{t}=x_1-(1-\sigma)x_0
\end{split}
\end{equation}

Similar to PitchCFM, the speaker embedding $s$, and refined $F_0$ embedding $f_r$ from PithCFM, are fed into UNet-based Mel-spectrogram score estimator for vector field prediction:
\begin{equation}
\begin{split}   
        v_{t}^{\mathrm{pred}}=\mathrm{NN}_{mel}(x_t,t;s,f_r)
\end{split}
\end{equation}

The training objectice is defined as:
\begin{equation}
\begin{split}   
        \mathcal{L}_{mel}=\| v_{t}^{pred}-v_t \|^{2}
\end{split}
\end{equation}
At inference, the speech token from source speaker is used as a prior. VoiceCFM generates the converted speech conditioned with target speaker style and the refined $F0$ from PitchCFM.

\begin{figure}[t!]
% \vspace{-0.5cm}
    \centering
    \includegraphics[width=8.0cm]{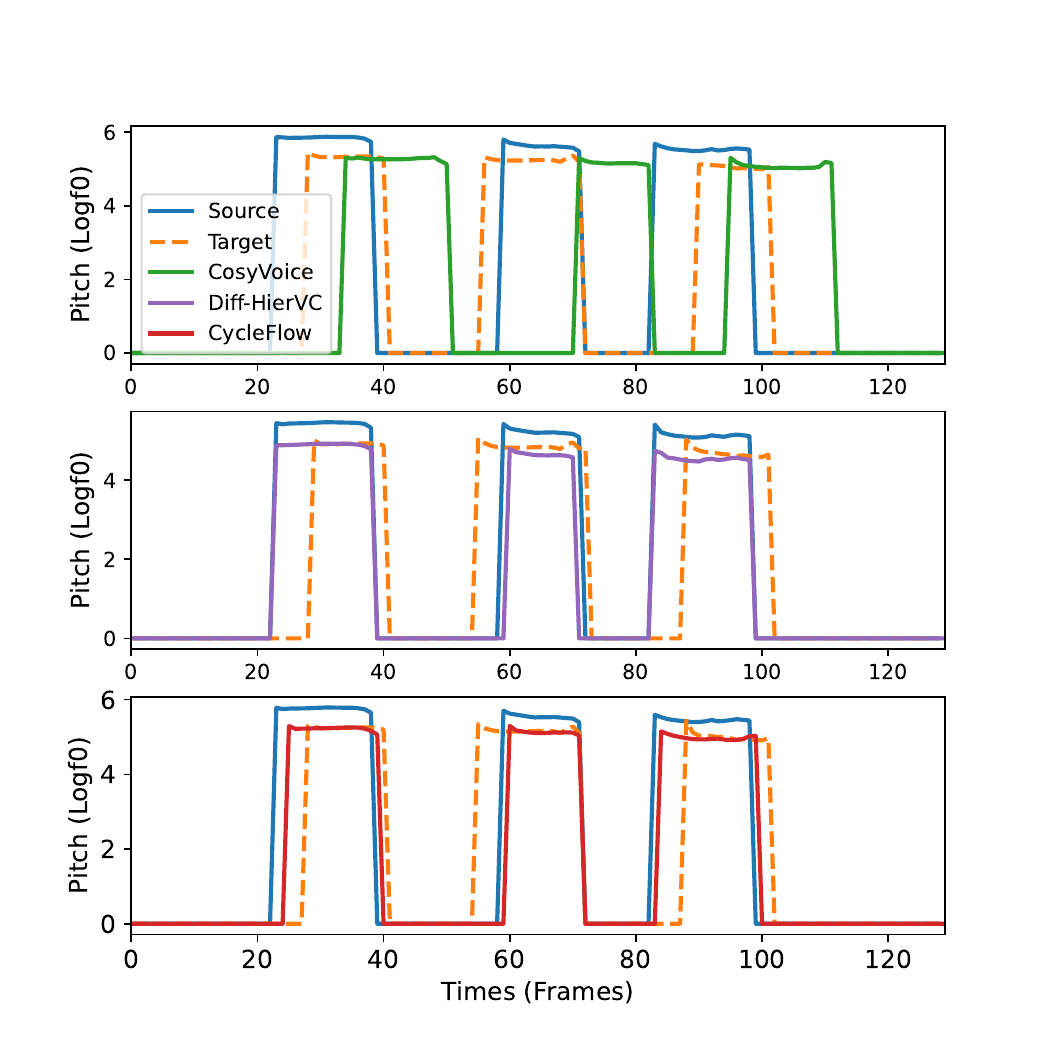}
    \vspace{-0.1cm}
    \caption{ Pitch contours without normalization of voice conversion results with the content 'Please call Stella'.} 
    % \vspace{-0.3cm}
    \label{pitchcontour}
\end{figure}

\begin{table*}[t!]
\caption{Intra-domain and cross-domain evaluation results of different methods with the confidence interval 95\%. MOS and SMOS mean subjective naturalness and speaker similarity, respectively. TSIM and WER mean timbre similarity calculated by \cite{WangZCC023} and word error rate. $F0$ PCC means the Pearson correlation coefficient between F0 of source and the converted audio.
}
% \vspace{-0.1cm}
\label{tab1}
\centering
% \small
\scalebox{1.0}{
% \begin{tabular}{@{}c|cc|cc@{}}
\begin{tabular}{ccccccccccc}
\toprule
\cmidrule(r){1-11}
\multirow{2.5}{*}{  \textbf{Methods}} & \multicolumn{5}{c}{\textbf{Intra-domain VC}} & \multicolumn{5}{c}{\textbf{Cross-domain VC}} \cr 
\cmidrule(lr){2-6} \cmidrule(lr){7-11} &\textbf{MOS$\uparrow$} &\textbf{SMOS$\uparrow$} &\textbf{TSIM$\uparrow$} &\textbf{$\log F0$ PCC$\uparrow$} &\textbf{WER$\downarrow$} &\textbf{MOS $\uparrow$} &\textbf{SMOS$\uparrow$} &\textbf{TSIM$\uparrow$} &\textbf{$\log F0$ PCC$\uparrow$} &\textbf{WER$\downarrow$}\cr   \midrule
DiffVC \cite{PopovVGSKW22} & 3.50$\pm$0.08   & 2.76$\pm$0.11  & 0.571 & 0.670 & 11.76\%   
& 3.22$\pm$0.06  & 2.58$\pm$0.09  & 0.514 & 0.594 & 13.70\%    \\
UUVC \cite{UUVC}    & 3.56$\pm$0.06    & 2.88$\pm$0.09  & 0.637 & 0.686 & 5.90\%
& 3.35$\pm$0.09   & 2.65$\pm$0.08  & 0.640 & 0.675 & 6.52\% \\
CosyVoice\cite{CosyVoice} & 3.65$\pm$0.08   & 3.03$\pm$0.09 & 0.785 & 0.732 & 3.51\%
& 3.44$\pm$0.07   & 2.89$\pm$0.08  & 0.759 & 0.725 & 4.18\% \\
Diff-HierVC \cite{Diff-HierVC} &3.60$\pm$0.09 & 3.15$\pm$0.11 & 0.741 & 0.785 & 3.58\%
& 3.49$\pm$0.06 & 3.03$\pm$0.07 & 0.705 & 0.746 & 4.20\% \\
% \midrule
\textbf{CycleFlow} &\bfseries 3.71$\pm$0.09 &\bfseries 3.23$\pm$0.06 &\bfseries 0.856 &\bfseries 0.813 &\bfseries 3.46\%
&\bfseries 3.52$\pm$0.06 &\bfseries 3.07$\pm$0.09 &\bfseries 0.822 &\bfseries 0.763 &\bfseries 4.04\% \\
\bottomrule
\end{tabular}}
% \vspace{-0.3cm}
\end{table*}

\section{Experiment}
\subsection{Experiment setup}
\para{Datasets and Configurations:}For intra-domain VC, we conduct our experiments using the LibriTTS dataset \cite{libritts}, which comprises 580 hours of speech from 2400 speakers.
Specifically, we select training sets of 500 male and 500 female speakers. 
It covers all inter-gender conversions: F2M and M2F, where F and M
indicate female and male respectively.
Since different genders lead to different distributions of pitch ranges, we construct an inter-gender VC to evaluate the effect of speaker pitch adaptation.
The samples of each speaker are randomly partitioned at a ratio of approximately 9:1 to obtain training and testing sets.
For cross-domain VC, we further utilize VCTK \cite{vctk} as unseen speaker style data, ensuring coverage across multiple vocal ranges.
The sampling rate of all speech datasets is set to 22.05kHz.

\para{Training:}
In our implementation, we employ the speech tokenizer with a single codebook of 4096 codes, following the approach outlined in \cite{CosyVoice}.
The 80-dimensional Mel-spectrogram was extracted with a frame size of 1024 and frame shift of 256. A pre-trained hifinet vocoder \cite{hifinet} is used to convert Mel-spectrogram back to the waveform.
Both PitchCFM and VoiceCFM adopt a UNet structure with 20 blocks, each having a dimension of 256.
We train CycleFlow for 80k iterations with 8 A100-80M GPUs, and set the learning rate to 1e-4 with warmup step 10,000.

\para{Baseline systems:}We compare our model with other systems. 1) DiffVC \cite{PopovVGSKW22}: The first work to introduce the diffusion probabilistic model into VC. 2) UUVC \cite{UUVC}: It applies self-supervised discrete representations for VC. 3) CosyVoice \cite{CosyVoice}: We leverage its CFM module to build a VC model based on supervised semantic tokens. 4) Diff-HierVC \cite{Diff-HierVC}: Diffusion-based hiverarchical VC for converted pitch and mel spectrogram generation.

\subsection{Analysis on F0 prediction}

Please note we use parallel speech data to visualize the results as shown in Fig.~\ref{pitchcontour}. 
For timbre conversion, the pitch contour of the converted speech matches average pitch of the target speech but retains detailed characteristics of the source pitch contour. 
Although the average pitch of CosyVoice matches the target speech, the rhythm is offset from the source speech. The rhythm of Diff-HierVC and CycleFlow is accurate, and the converted pitch contour is located at the source speech position as expected.
However, CycleFlow matches the target speech more closely in terms of average pitch, attributing to the pitch refinement of VoiceCFM.

\subsection{Comparison of VC tasks}
To assess the effect of our method, we conducted subjective and objective evaluation on different methods. 10 listeners (5 males and 5 females) participated in subjective evaluation and scored the naturalness (MOS) and similarity (SMOS) of the test set from 1-5 points.
In addition to the performance in the general intra-domain VC, we also assess the performance in the cross-domain VC with differences between source and target speaker's vocal ranges or in cases of wide-ranging pitch shifting. We randomly select 25 samples from test set to calculate scores.
As shown in Tab~\ref{tab1}, CycleFlow outperforms the baseline models in all metrics.
The significant improvement in TSIM and $F0$ PCC metrics demonstrates that the introduction of cycle consistency regularization and PitchCFM can improve the adaptation of speaker timbre and pitch respectively.

\subsection{Evaluation of speaker similarity}
\begin{figure}[h!]
% \vspace{-0.3cm}
    \centering
    \includegraphics[width=8.8cm]{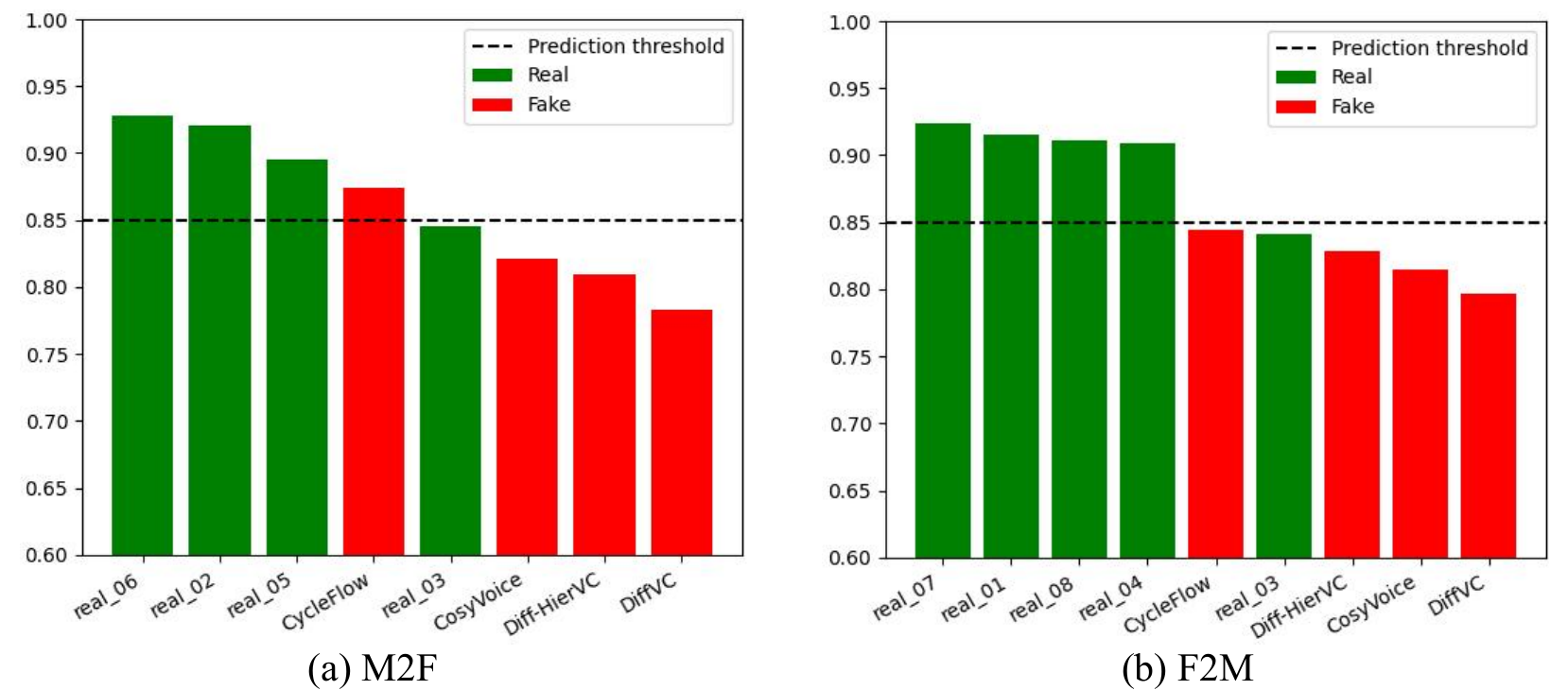}
    % \vspace{-0.1cm}
    \caption{Objective evaluation results for VC. F: Female; M: Male. Green groups are real speech. Red groups are synthesized speech from different models.} 
    % \vspace{-0.3cm}
    \label{FSD}
\end{figure}

To conduct further objective evaluation of different models, we apply an open-source speech fake detection toolkit\footnote{Resemblyzer toolkit: https://github.com/resemble-ai/Resemblyzer} to evalute the speaker similarity. 
To compare the speaker timbre similarity between converted voice and real voice, it will give a score ranging from 0 to 1. A higher score signifies a greater similar between the fake voice and the real voice. 
As shown in Fig.~\ref{FSD}, the dashed line indicates the score that passes the toolkit test. CycleFlow outperforms other baselines in speaker similarity.

\subsection{Ablation study}
As shown in Tab~\ref{ablation}, we train our model with specific components selectively removed to evaluate the effect of the proposed components.
The significant decrease of $\mathcal{L}_{cycle}$ on SMOS and TSIM proves the positive effect of the cycle consistency regularization on speaker timbre similarity. The lack of PitchCFM leads to a significant drop in $\log F0$ PCC, indicating that it plays a crucial role in the pitch correction of cross-domain VC.

\begin{table}[t!]
\centering
\caption{Ablation studies on cross-domain VC task with unseen speakers from VCTK dataset.}
% \vspace{-0.1cm}
\label{ablation}
\resizebox{1.0\columnwidth}{!}{
\begin{tabular}{cccccc}
\toprule
\textbf{Method} & \textbf{MOS$\uparrow$} & \textbf{SMOS$\uparrow$}  & \textbf{TSIM$\uparrow$} &\textbf{$\log F0$} \textbf{PCC$\uparrow$} &\textbf{WER$\downarrow$}  \\
\midrule
\textbf{Our} & /  & / & /  & /  & / \\
\textbf{w/o} $\mathcal{L}_{cycle}$   & -0.12 & -0.20 & -0.11 &  -0.03  & -0.05\% \\
\textbf{w/o} PitchCFM  & -0.05 & -0.16 &  -0.14  & -0.10  & / \\
\bottomrule
\end{tabular}
}
\vspace{-0.1cm}
\end{table}

\section{Conclusion}
In this paper, We propose a novel VC called CycleFlow, leveraging cycle consistency in condition flow matching for speaker style adaptation.
To bridge the train-inference mismatch in non-parallel VC and improve speaker timbre similarity, we design cycle consistency regularization to learn the mapping between source and target speaker style domains. Additionally, we design a Dual-CFM to alleviate the hoarseness in cross-domain VC, where the pitch span is large between source domain speaker and target domain speaker.
Experiments demonstrate CycleFlow's superiority in both speaekr timbre and pitch adaptation in voice conversion.

\section{Acknowledgement}
This paper is supported by the Key Research and Development Program of Guangdong Province under grant No.2021B0101400003. Corresponding author is Jianzong Wang from Ping An Technology (Shenzhen) Co., Ltd. (jzwang@188.com).

% \vfill\pagebreak

\bibliographystyle{IEEEbib}
\bibliography{CycleFlow}

\begin{thebibliography}{10}

\bibitem{UUVC}
Li{-}Wei Chen, Shinji Watanabe, and Alexander Rudnicky,
\newblock ``A unified one-shot prosody and speaker conversion system with self-supervised discrete speech units,''
\newblock in {\em {IEEE} International Conference on Acoustics, Speech and Signal Processing {ICASSP} 2023, Rhodes Island, Greece, June 4-10, 2023}, 2023, pp. 1--5.

\bibitem{PolyakACKLHMD21}
Adam Polyak, Yossi Adi, Jade Copet, Eugene Kharitonov, Kushal Lakhotia, Wei{-}Ning Hsu, Abdelrahman Mohamed, and Emmanuel Dupoux,
\newblock ``Speech resynthesis from discrete disentangled self-supervised representations,''
\newblock in {\em Interspeech}, 2021, pp. 3615--3619.

\bibitem{eadvc}
Ziqi Liang, Jianzong Wang, Xulong Zhang, Yong Zhang, Ning Cheng, and Jing Xiao,
\newblock ``{EAD-VC:} enhancing speech auto-disentanglement for voice conversion with {IFUB} estimator and joint text-guided consistent learning,''
\newblock in {\em International Joint Conference on Neural Networks, {IJCNN} 2024, Yokohama, Japan, June 30 - July 5, 2024}, 2024, pp. 1--7.

\bibitem{NANSY}
Hyeong{-}Seok Choi, Juheon Lee, Wansoo Kim, Jie Lee, Hoon Heo, and Kyogu Lee,
\newblock ``Neural analysis and synthesis: Reconstructing speech from self-supervised representations,''
\newblock in {\em Advances in Neural Information Processing Systems 34: Annual Conference on Neural Information Processing Systems 2021, NeurIPS 2021, December 6-14, 2021, virtual}, 2021, pp. 16251--16265.

\bibitem{DengT0W0023}
Yimin Deng, Huaizhen Tang, Xulong Zhang, Jianzong Wang, Ning Cheng, and Jing Xiao,
\newblock ``{PMVC:} data augmentation-based prosody modeling for expressive voice conversion,''
\newblock in {\em Proceedings of the 31st {ACM} International Conference on Multimedia, {MM} 2023, Ottawa, ON, Canada, 29 October 2023- 3 November 2023}, 2023, pp. 184--192.

\bibitem{wav2vec2}
Alexei Baevski, Yuhao Zhou, Abdelrahman Mohamed, and Michael Auli,
\newblock ``wav2vec 2.0: {A} framework for self-supervised learning of speech representations,''
\newblock in {\em Advances in Neural Information Processing Systems 33: Annual Conference on Neural Information Processing Systems 2020, NeurIPS 2020, December 6-12, 2020, virtual}, 2020.

\bibitem{hubert}
Wei{-}Ning Hsu, Benjamin Bolte, Yao{-}Hung~Hubert Tsai, Kushal Lakhotia, Ruslan Salakhutdinov, and Abdelrahman Mohamed,
\newblock ``Hubert: Self-supervised speech representation learning by masked prediction of hidden units,''
\newblock {\em {IEEE} {ACM} Trans. Audio Speech Lang. Process.}, vol. 29, pp. 3451--3460, 2021.

\bibitem{contentvec}
Kaizhi Qian, Yang Zhang, Heting Gao, Junrui Ni, Cheng{-}I Lai, David~D. Cox, Mark Hasegawa{-}Johnson, and Shiyu Chang,
\newblock ``Contentvec: An improved self-supervised speech representation by disentangling speakers,''
\newblock in {\em International Conference on Machine Learning, {ICML} 2022, 17-23 July 2022, Baltimore, Maryland, {USA}}, 2022, vol. 162 of {\em Proceedings of Machine Learning Research}, pp. 18003--18017.

\bibitem{CosyVoice}
Zhihao Du, Qian Chen, Shiliang Zhang, Kai Hu, Heng Lu, Yexin Yang, Hangrui Hu, Siqi Zheng, Yue Gu, Ziyang Ma, Zhifu Gao, and Zhijie Yan,
\newblock ``Cosyvoice: {A} scalable multilingual zero-shot text-to-speech synthesizer based on supervised semantic tokens,''
\newblock {\em CoRR}, vol. abs/2407.05407, 2024.

\bibitem{soundstorm}
Zal{\'{a}}n Borsos, Matthew Sharifi, Damien Vincent, Eugene Kharitonov, Neil Zeghidour, and Marco Tagliasacchi,
\newblock ``Soundstorm: Efficient parallel audio generation,''
\newblock {\em CoRR}, vol. abs/2305.09636, 2023.

\bibitem{SEF_VC}
Junjie Li, Yiwei Guo, Xie Chen, and Kai Yu,
\newblock ``Sef-vc: Speaker embedding free zero-shot voice conversion with cross attention,''
\newblock in {\em ICASSP}, 2024, pp. 12296--12300.

\bibitem{WangCXTW23}
Zhichao Wang, Yuanzhe Chen, Lei Xie, Qiao Tian, and Yuping Wang,
\newblock ``{LM-VC:} zero-shot voice conversion via speech generation based on language models,''
\newblock {\em {IEEE} Signal Process. Lett.}, vol. 30, pp. 1157--1161, 2023.

\bibitem{TangZWCX23a}
Huaizhen Tang, Xulong Zhang, Jianzong Wang, Ning Cheng, and Jing Xiao,
\newblock ``{VQ-CL:} learning disentangled speech representations with contrastive learning and vector quantization,''
\newblock in {\em {IEEE} International Conference on Acoustics, Speech and Signal Processing {ICASSP} 2023, Rhodes Island, Greece, June 4-10, 2023}, 2023, pp. 1--5.

\bibitem{cycleganvc}
Fuming Fang, Junichi Yamagishi, Isao Echizen, and Jaime Lorenzo{-}Trueba,
\newblock ``High-quality nonparallel voice conversion based on cycle-consistent adversarial network,''
\newblock in {\em 2018 {IEEE} International Conference on Acoustics, Speech and Signal Processing, {ICASSP} 2018, Calgary, AB, Canada, April 15-20, 2018}, 2018, pp. 5279--5283.

\bibitem{cycleganvc2}
Takuhiro Kaneko, Hirokazu Kameoka, Kou Tanaka, and Nobukatsu Hojo,
\newblock ``Cyclegan-vc2: Improved cyclegan-based non-parallel voice conversion,''
\newblock in {\em {IEEE} International Conference on Acoustics, Speech and Signal Processing, {ICASSP} 2019, Brighton, United Kingdom, May 12-17, 2019}, 2019, pp. 6820--6824.

\bibitem{maskcycleganvc}
Takuhiro Kaneko, Hirokazu Kameoka, Kou Tanaka, and Nobukatsu Hojo,
\newblock ``Maskcyclegan-vc: Learning non-parallel voice conversion with filling in frames,''
\newblock in {\em {IEEE} International Conference on Acoustics, Speech and Signal Processing, {ICASSP} 2021, Toronto, ON, Canada, June 6-11, 2021}, 2021, pp. 5919--5923.

\bibitem{ZhangWCXX21}
Xulong Zhang, Jianzong Wang, Ning Cheng, Edward Xiao, and Jing Xiao,
\newblock ``Cyclegean: Cycle generative enhanced adversarial network for voice conversion,''
\newblock in {\em {IEEE} Automatic Speech Recognition and Understanding Workshop, {ASRU} 2021, Cartagena, Colombia, December 13-17, 2021}, 2021, pp. 930--937.

\bibitem{WeiCDC23}
Haojie Wei, Xueke Cao, Tangpeng Dan, and Yueguo Chen,
\newblock ``{RMVPE:} {A} robust model for vocal pitch estimation in polyphonic music,''
\newblock in {\em 24th Annual Conference of the International Speech Communication Association, Interspeech 2023, Dublin, Ireland, August 20-24, 2023}, 2023, pp. 5421--5425.

\bibitem{WangZCC023}
Hui Wang, Siqi Zheng, Yafeng Chen, Luyao Cheng, and Qian Chen,
\newblock ``{CAM++:} {A} fast and efficient network for speaker verification using context-aware masking,''
\newblock in {\em 24th Annual Conference of the International Speech Communication Association, Interspeech 2023, Dublin, Ireland, August 20-24, 2023}, 2023, pp. 5301--5305.

\bibitem{PopovVGSKW22}
Vadim Popov, Ivan Vovk, Vladimir Gogoryan, Tasnima Sadekova, Mikhail~Sergeevich Kudinov, and Jiansheng Wei,
\newblock ``Diffusion-based voice conversion with fast maximum likelihood sampling scheme,''
\newblock in {\em The Tenth International Conference on Learning Representations, {ICLR} 2022, Virtual Event, April 25-29, 2022}, 2022.

\bibitem{Diff-HierVC}
Ha{-}Yeong Choi, Sang{-}Hoon Lee, and Seong{-}Whan Lee,
\newblock ``Diff-hiervc: Diffusion-based hierarchical voice conversion with robust pitch generation and masked prior for zero-shot speaker adaptation,''
\newblock in {\em 24th Annual Conference of the International Speech Communication Association, Interspeech 2023, Dublin, Ireland, August 20-24, 2023}, 2023, pp. 2283--2287.

\bibitem{ns2}
Kai Shen, Zeqian Ju, Xu~Tan, Eric Liu, Yichong Leng, Lei He, Tao Qin, Sheng Zhao, and Jiang Bian,
\newblock ``Naturalspeech 2: Latent diffusion models are natural and zero-shot speech and singing synthesizers,''
\newblock in {\em The Twelfth International Conference on Learning Representations, {ICLR} 2024, Vienna, Austria, May 7-11, 2024}, 2024.

\bibitem{LipmanCBNL23}
Yaron Lipman, Ricky T.~Q. Chen, Heli Ben{-}Hamu, Maximilian Nickel, and Matthew Le,
\newblock ``Flow matching for generative modeling,''
\newblock in {\em The Eleventh International Conference on Learning Representations, {ICLR} 2023, Kigali, Rwanda, May 1-5, 2023}, 2023.

\bibitem{0001FMHZRWB24}
Alexander Tong, Kilian Fatras, Nikolay Malkin, Guillaume Huguet, Yanlei Zhang, Jarrid Rector{-}Brooks, Guy Wolf, and Yoshua Bengio,
\newblock ``Improving and generalizing flow-based generative models with minibatch optimal transport,''
\newblock {\em Trans. Mach. Learn. Res.}, vol. 2024, 2024.

\bibitem{conformer}
Anmol Gulati, James Qin, Chung{-}Cheng Chiu, Niki Parmar, Yu~Zhang, Jiahui Yu, Wei Han, Shibo Wang, Zhengdong Zhang, Yonghui Wu, and Ruoming Pang,
\newblock ``Conformer: Convolution-augmented transformer for speech recognition,''
\newblock in {\em 21st Annual Conference of the International Speech Communication Association, Interspeech 2020, Virtual Event, Shanghai, China, October 25-29, 2020}, 2020, pp. 5036--5040.

\bibitem{libritts}
Heiga Zen, Viet Dang, Rob Clark, Yu~Zhang, Ron~J. Weiss, Ye~Jia, Zhifeng Chen, and Yonghui Wu,
\newblock ``Libritts: {A} corpus derived from librispeech for text-to-speech,''
\newblock in {\em 20th Annual Conference of the International Speech Communication Association, Interspeech 2019, Graz, Austria, September 15-19, 2019}, 2019, pp. 1526--1530.

\bibitem{vctk}
Christophe Veaux, Junichi Yamagishi, and Kirsten MacDonald,
\newblock ``Superseded - cstr vctk corpus: English multi-speaker corpus for cstr voice cloning toolkit,''
\newblock 2016.

\bibitem{hifinet}
Yinghao~Aaron Li, Cong Han, Xilin Jiang, and Nima Mesgarani,
\newblock ``Hiftnet: {A} fast high-quality neural vocoder with harmonic-plus-noise filter and inverse short time fourier transform,''
\newblock {\em CoRR}, vol. abs/2309.09493, 2023.

\end{thebibliography}

\end{document}